\documentclass[psfig,preprint]{emulateapj}
\usepackage{apjfonts}
\usepackage{graphicx}
\usepackage{natbib}
\newcommand{\etal}{{et al.} }
\newcommand{\rxte}{{\it RXTE} }

\newcommand{\asca}{{\it ASCA} }

\newcommand{\xmmp}{{\it XMM-Newton}}

\newcommand{\feka}{{Fe~K$\alpha$} }
\newcommand{\fekap}{{Fe~K$\alpha$}}
\newcommand{\fekb}{{Fe~K$\beta$} }

\newcommand{\gp}{{$g_{+}$} }
\newcommand{\gpp}{{$g_{+}$}}
\newcommand{\gm}{{$g_{-}$} }
\newcommand{\gmp}{{$g_{-}$}}
\newcommand{\am}{{$a/M$} }
\newcommand{\amp}{{$a/M$}}
\newcommand{\rhorz}{{$R_{\rm H}$} }
\newcommand{\rhorzp}{{$R_{\rm H}$}}
\newcommand{\rms}{{$R_{\rm MS}$} }
\newcommand{\rmsp}{{$R_{\rm MS}$}}
\newcommand{\rd}{{$R_{\rm d}$} }
\newcommand{\rdp}{{$R_{\rm d}$}}
\newcommand{\rg}{{$R_{\rm G}$} }
\newcommand{\rgp}{{$R_{\rm G}$}}

\newcommand{\thetaobs}{{$\theta_{\rm obs}$} }
\newcommand{\thetaobsp}{{$\theta_{\rm obs}$}}

\newcommand{\figrhrmsbegin}{{Fig.~1} }
\newcommand{\figtorbit}{{Fig.~2} }

\newcommand{\figprofile}{{Fig.~3} }

\newcommand{\figavsr}{{Fig.~4} }
\newcommand{\figavsrp}{{Fig.~4}}
\newcommand{\figavsrerr}{{Fig.~5} }
\newcommand{\figavsrerrp}{{Fig.~5}}

\newcommand{\figthetaerrp}{{Fig.~6}}
\newcommand{\figavsrexamps}{{Fig.~7} }
\newcommand{\figavsrexampsp}{{Fig.~7}}

\newcommand{\figavsrexampssp}{{Figs.~7}}

\begin{document}
\title{On the Prospect of Constraining 
Black-Hole Spin Through
X-ray Spectroscopy of Hotspots}
\author{Kendrah D. Murphy}\altaffilmark{1}
\affil{MIT Kavli Institute for Astrophysics and Space Research, 77 Massachusetts
Avenue, NE 80, Cambridge, MA 02139.}
\author{Tahir Yaqoob}\altaffilmark{2}
\affil{Department of Physics and Astronomy,
Johns Hopkins University, Baltimore, MD 21218.}
\author{Michal Dov\v{c}iak}
\author{Vladimir Karas}
\affil{Astronomical Institute, Academy of Sciences, Bo\v{c}n\'{\i} II, CZ-141 31 Prague, Czech Republic.}

\altaffiltext{1}{Department of Physics and Astronomy,
Johns Hopkins University, Baltimore, MD 21218.}
\altaffiltext{2}{Astrophysics Science Division,
NASA/Goddard Space Flight Center, Greenbelt, MD 20771.}

\date{Received /Accepted }

\begin{abstract}

Future X-ray instrumentation is expected to allow us to significantly improve the constraints derived
from the Fe~K lines in AGN, such as the black-hole angular momentum (spin) and the inclination angle of
the putative accretion disk. 
We consider the possibility that measurements of the persistent,
time-averaged Fe~K line emission from the disk could be supplemented by the observation of a localized
flare, or ``hotspot'', orbiting close to the black hole. 
Although observationally challenging, such measurements would recover some of the information loss that is
inherent to the radially-integrated line profiles.  We present calculations for this scenario to assess
the extent to which, in principle, black-hole spin may be measured. We quantify the feasibility of this
approach using realistic assumptions about likely measurement uncertainties.

\end{abstract}

\keywords{black hole physics - galaxies: active - line: profiles - X-ray: galaxies}

\section{Introduction} 
\label{introavsr}

Black holes are often described as having ``no hair" as there are only three characteristic quantities that
define them: mass,  charge\footnote{This may include magnetic charge in addition to electric charge.},  and
angular momentum.  Evaluating the black-hole metric (``Kerr" metric, or when charge is non-zero,
``Kerr-Newman" metric) amounts to accurate measurement of these quantities.  Ultimately these parameters
will be critical for testing general relativity.  
The masses of black holes are measurable and there
are continual advances in mass-measurement methods (e.g. Peterson \& Benz 2008, and references therein).  As
yet, however, a method for measuring black-hole charge has not been discovered.  The electric  charge is
usually assumed to be zero since astronomical black holes are embedded in plasma, so the  process of
selective charge accretion gradually diminishes the charge to insignificant levels.  It is therefore
essential to determine the angular momentum (spin) of a black hole in order to constrain its metric.  In
addition, the spin of the supermassive black hole found in an AGN is an important diagnostic of the system. 
For example, black-hole spin is thought to power relativistic jets seen in some AGN sources (see Blandford
\& Znajek 1977).  Furthermore, the growth history of supermassive black holes may involve spin evolution,
ultimately affecting the present distribution of spins (e.g.  King \etal 2008).  

Accretion from a disk is thought to change the spin of a black hole until an equilibrium value is
established. The spin can be expressed in the 
dimensionless geometrical units \am (where $M$ is the mass of the black
hole), which has an absolute value between 0 (non-spinning, Schwarzschild black hole) and 1 (extreme Kerr
black hole). It is suggested that the upper limit on spin due to accretion is \am$\sim0.9982$ (Thorne 1974),
although this value has been challenged (e.g. Gammie \etal 2004; Beckwith \etal 2008). 
The influence of a rotating
gravitational field on light is specific to general relativity. It is not reproduced in Newtonian
theory (e.g. Islam 1985). The spin of a black
hole determines the horizon radius (\rhorzp) and the radius of the marginally stable orbit (\rmsp), otherwise
known as the innermost stable circular orbit (ISCO), two quantities that are predicted by general
relativity.  \figrhrmsbegin shows the dependence of \rhorz and \rms (as functions of gravitational radius,
\rg$\equiv GM/c^{2}$; see e.g. Misner \etal 1973) on spin.  The ISCO of a non-spinning black
hole is located at 6\rg from the center, but this radius shrinks to 1.227 \rg  (1.0 \rgp) for a black hole
with \am$=0.9982 \ (1.0)$. 

\begin{figure}[!htb]
\begin{center}
	\epsscale{1.0}
        \plotone{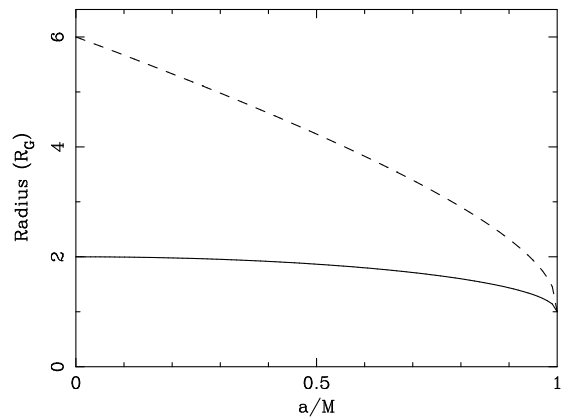}
\end{center}
\caption{\footnotesize Distance (in gravitational radii) from the black hole to the
	horizon radius (\rhorzp; solid) and to the marginally
	stable orbit (\rmsp; dashed), versus spin.  See Misner
	\etal (1973).}
\end{figure}

Since we are unable to image the vicinity of the supermassive black hole in AGNs, the spectrum of the
X-ray emission, which originates from the innermost regions of the system, is one of the best probes
that we currently have of the strong-gravity regime of these systems. The Fe~K emission line in the X-ray
spectrum is also often used as a tool to probe the physical parameters of an AGN system.  Narrow Fe~K
lines that are not relativistically shifted in energy, which are ubiquitous in AGN observations (see
Yaqoob \& Padmanabhan 2004), are most likely reprocessed emission from circumnuclear matter far from
the central black hole and/or from the outer regions of the accretion disk.  On the other hand, broad
Fe~K lines, which are detected less often, are most likely the product of emission that originates in
the inner accretion disk (e.g. Nandra \etal 2007).  This emission, which is subject to gravitational
and relativistic Doppler shifting in energy, is therefore believed to uncover important information
on the innermost regions of AGNs, including the spin of the black hole
(e.g. see Brenneman \& Reynolds 2006; Miller 2007, and references therein).
Similar techniques have been applied to stellar-mass black holes in X-ray binary systems
(e.g. Miller \etal 2008; Reis \etal 2009; Miller \etal 2009, \& references therein). 
In addition to the relativistic Fe~K emission-line, X-ray binaries
(and in principle AGN as well) offer the possibility of constraining black-hole spin from
spectral-fitting of the accretion-disk continuum, which is influenced by
relativistic effects in the strong gravity regime (e.g. see Shafee \etal
2006; McClintock \etal 2006; Miller \etal 2009, \& references therein).
It has also been suggested that black-hole spin
measurements in Galactic X-ray binary systems may be
possible through timing measurements of the high-frequency quasi-periodic
oscillations (QPOs; e.g. McClintock \& Remillard 2006). 

In the present paper we describe a method to constrain black-hole spin
that involves measurements of Fe~K lines resulting from localized,
accretion-disk flares, or ``hotspots'' that orbit the black hole. The method
relies on the measurement of the energies of narrow spectral
features, and not line intensities. Some of the physical information
that is lost in the radially-integrated Fe~K emission-line profiles
is recovered, since the spatial scale that is probed by the hotspots
is much smaller. In \S\ref{hotspot} we give a detailed description of
how Fe~K line hotspots may be used to constrain black-hole spin
and investigate the impact of likely observational uncertainties
on the proposed method. The results of the investigation of the 
technique are described in
\S\ref{avsrresults}. Our conclusions are summarized in \S\ref{conclusions}.

\section{Measuring Black-Hole Spin From Localized Hotspots}
\label{hotspot}

It is theorized that X-ray flares can result from magnetic reconnection events that illuminate
localized portions of the accretion disk, where the emission is reprocessed (e.g. Galeev \etal 1979;
Czerny \etal 2004, and references therein).  These compact ``hotspots" temporarily enhance the
continuum and Fe~K line emission from the localized region of the accretion disk. The enhanced Fe~K
line emission contains supplementary information to that of the persistent, radially-integrated
emission that is more often observed and measured.  In theory, such
hotspots may co-rotate with the accretion disk.  The orbital time-scale depends on the mass and spin of
the central black hole and the distance from the black hole to the orbiting spot (e.g. Bardeen \etal
1972; Goosmann \etal 2006).  In \figtorbit we show calculations of the orbital time versus distance
from the innermost stable circular orbit for a black hole with a mass of $10^{7} M_{\odot}$ for five
different values of \am between 0 and 1.  The orbital time can simply be scaled linearly for other
masses.  In the limit of large $R$, the curves converge to the simple Newtonian relation, $t_{\rm
orbit} \sim 100 \pi R^{1.5}(M/10^{7} M_{\odot})$ s, where $R$ is in units of \rgp.

\begin{figure}[!htb]
\begin{center}
       \epsscale{1.0}
       \plotone{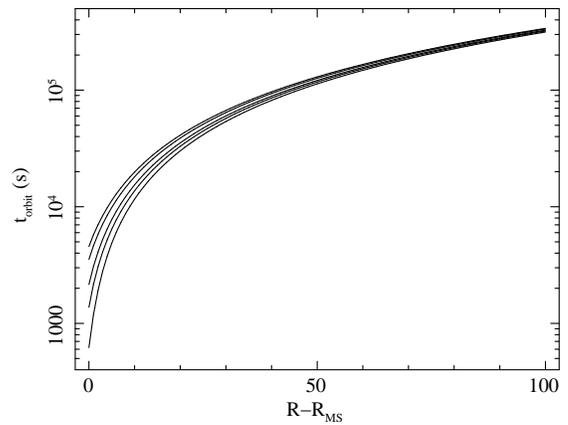}
\end{center}
       \caption{ \footnotesize Orbital time versus distance from the innermost stable circular
       orbit, in units of gravitational radii, for a black hole with a mass of
       $10^{7} M_{\odot}$.  The orbital time can simply be scaled linearly for other masses.  
       Curves are shown (top to bottom) for \am $=0,\ 0.3, \ 0.7, \ 0.9,$ and 1.}
\end{figure}


A number of AGNs show observational evidence of emission from localized hotspots that originates in
the inner regions of the accretion disk (e.g. Iwasawa \etal 1999; Turner \etal 2002; Guainazzi 2003;
Iwasawa \etal 2004; Turner \etal 2006; Murphy \etal 2007).  For example, during a 
flare seen in the \asca data for MCG~-6-30-15, the centroid of the Fe~K line emission was observed to
redshift from 6.4 keV to below 6 keV, implying that the profile was dominated by localized emission
close to the black hole (Iwasawa \etal 1999).  Similar behavior was found from the \rxte data of
NGC~2992 (see Murphy \etal 2007).  Theoretically, with long observations yielding
high-resolution, high-throughput spectral data we will be able to detect full (possibly multiple)
orbits of accretion-disk hotspots, depending on the black-hole mass.  In fact, Iwasawa \etal (2004)
claim to have already observed four full, consecutive orbits of a hotspot around the black hole in
NGC~3516 with \xmmp.  

\subsection{Method for Constraining Black-Hole Spin}
\label{hotspotmeth}

We develop the method that was suggested in Yaqoob (2001) for using the Fe~K line resulting
from a hotspot to constrain black-hole spin.  
Properties of the Fe~K line emission from hotspots
have been discussed extensively
in the literature 
(e.g. Nayakshin \& Kazanas 2001;
Goosmann et al. 2007; Dov\v{c}iak et al. 2008, and references therein).
Here we formalize an approach to specifically use such emission
to measure black-hole spin.
We emphasize that this work cannot yet be applied to
current data since the available X-ray instrumentation does not have the required energy resolution
and effective area.  

In the remainder of this section we describe theoretical results pertaining to the Fe~K line that make use
of tables of transfer function\footnote{The usage of the term ``transfer function'' in the present paper 
(see Dov\v{c}iak \etal 2004 for a detailed description)
does not include the time domain and should not be confused with the usage of the term in Yaqoob (2001) which
does refer to the time domain.} 
calculations that have been created for the non-axisymmetric {\tt kyr1line}
model (Dov\v{c}iak \etal 2004). This model is included in a suite known as the ``{\tt KY} models"
(Dov\v{c}iak \etal 2004) that are available for analyzing relativistic X-ray line profiles from  black-hole
accretion disks in the Kerr metric.  The routine {\tt kyr1line} in particular models the instantaneous 
Fe~K line profile function from the accretion disk for a given distance from the spin-dependent horizon
radius (\rd$=R-$\rhorz), inclination angle of the disk with respect to the observer (\thetaobsp), and
azimuthal angle ($\phi$) on the disk.  We remind the reader that the model which we use here does not
assume an axially-symmetric accretion disk. Instead, localized spots can be tracked on the surface of the
disk, taking all light-bending and time-delay effects into account.

Assuming that a hotspot completes at least one full orbit, tracing out an annulus around the disk,
the resulting (temporally-integrated) Fe~K line emission will have a characteristic ``double-horned"
profile, with peaks corresponding to the extreme redshifts and blueshifts (due to Doppler and extreme
gravitational effects) of the hotspot emission relative to the rest energy of the line.  The more
localized the hotspot is (that is, the narrower the annulus is), the sharper the two peaks of the
profile will be. 

In \figprofile we show an example of a theoretical Fe~line profile function, which was produced with the
{\tt kyr1line} model.  The emission in this example is calculated at a single radius of $R=6$\rg from the
black hole for a disk with an inclination angle of \thetaobs$=60^{\circ}$.  We define the inclination angle
with respect to the axis of the accretion disk, where \thetaobs$=0^{\circ}$ corresponds to a face-on
observing angle and \thetaobs$=90^{\circ}$ corresponds to an edge-on observing angle.  
An inclination angle of $60^{\circ}$ corresponds to the average of a distribution of 
randomly oriented disks. For inclination angles greater than $60^{\circ}$,
disk-reflection features begin to diminish rapidly as the inclination angle increases (e.g. see
George \& Fabian 1991), and in AGN, heavy line-of-sight obscuration may become problematic
for inclination angles greater than $60^{\circ}$.
The horizontal axis
in \figprofile is given in units of the energy-shift factor, $g \equiv E_{\rm obs}/E_{0}$, where $E_{\rm
obs}$ is the observed energy at infinity and $E_{0}$ is the rest energy of the Fe~K line from the emitting
material.  We refer to the $g$ values of the red and blue extrema, for a given angular distribution of
photon emission, as \gm and \gpp, respectively. \figprofile shows \gm and \gp for isotropic emission in the
disk frame, in which case \gm and \gp correspond to the two peaks of the profile.  

\begin{figure}[!htb]
\begin{center}
       \epsscale{1.0}
       \plotone{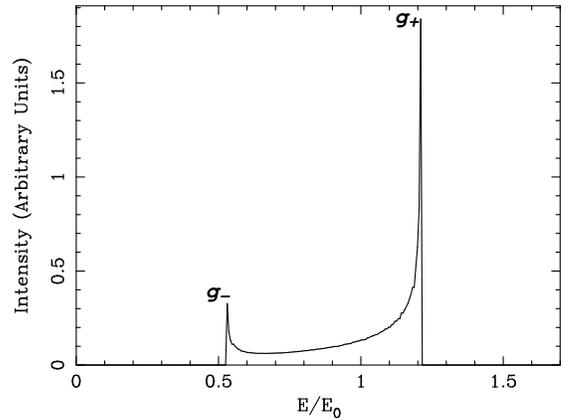}
\end{center}
       \caption{ \footnotesize Example of an Fe~K line profile function from the {\tt kyr1line}
       model for \thetaobs $= 60^{\circ}$ and \am $= 0$, at a single radius of 
       $R=$\rms$\equiv6$\rg from the black hole.  The $x$-axis is shown in units of
       energy-shift factor ($g$).  In these units, the redshifted and blueshifted extrema 
       may be defined as
       \gm and \gpp, respectively.}
\end{figure}

Given an angular distribution of photon emission from the orbiting hotspot co-rotating in the disk frame,
the values of \gm and \gp for the Fe~K line profile of an orbiting hotspot depend only on the radial
distance from the center of the black hole to the hotspot ($R$), the inclination angle of the accretion disk
(\thetaobsp), and the black hole spin ($a/M$).  In the present work we use transfer functions that assume
isotropic emission in the disk frame.  We therefore refer to \gm and \gp synonymously with the $g$ values
of the peaks of the
line profile.  However, the code allows one to specify a more complicated profile of limb-darkening or
limb-brightening (see Dov\v{c}iak \etal 2004).  For the moment we neglect the effects of the finite size of
the hotspot and calculate the best possible constraints before gradually introducing practical
uncertainties.  

It may be possible to obtain robust measurements of \thetaobs independently by modeling the
time-averaged, radially-integrated Fe~K line profile (that is, the emission across the 
entire accretion disk). The inclination angle is mainly affected by the blue-wing cut-off
of the Fe~K line profile, and is not sensitive to the other disk parameters or black-hole
spin. This is because the maximum blueshift is due to the Doppler-shift
maximum corresponding to emission from a particular radius of the disk.  Emission from inside this
particular radius is gravitationally redshifted away from the blue-wing cut-off, and outside this
radius the Doppler blueshift decreases for a given \thetaobs because of the lower Keplerian
velocities at larger radii.  If the outer radius of emission is larger than the radius that produces
the largest blueshift (as, from observation, generally appears to be the case), the blue wing is
then mostly affected by \thetaobsp.  Changes in the
inner disk radius or in the steepness of the radial emissivity
function mainly affect the red wing and not the blue
cut-off because the entire Doppler profiles from the inner radii are gravitationally redshifted.
Thus, if \thetaobs can be measured in this way,
measurements of \gm and \gp from a hotspot emission-line profile should then give constraints on the
distance to the hotspot ($R$) and the black-hole spin ($a/M$) since they are associated with two equations
for two unknown quantities. This, of course, assumes that we know the ionization state of Fe, since $g$
depends on the rest-frame energy of the emission line.  We address this uncertainty in
\S\ref{uncertainties}. 

Values of \gm and \gpp, for discrete values of $R$, \thetaobsp, and \amp, may be extracted from each of the
line-profile functions that are stored in the {\tt kyr1line} model tables.  Once these values of \gm and \gp
have been compiled, they can then be interpolated (in each of the three variable directions) to construct
curves of $a/M$ versus $R$ for a given inclination angle and any chosen constant value of \gm and \gpp. 
Thus, one can construct \am versus $R$ contours for any measured pair of \gm and \gp values when \thetaobs
is already known.  For a general (non-isotropic) angular distribution of emission, when the
peaks of the line profile do not occur at \gm and \gpp, the observed $g$ values of the profile peaks may be
calculated and related to \am and $R$ instead of \gm and \gpp.  

In \figavsrp, we illustrate how this information can potentially determine constraints on \am and
$R$ from a temporally-integrated Fe~K line profile from a complete orbit of a localized
accretion-disk hotspot.  In this example, we assume that \thetaobs has been independently measured
to be $60^{\circ}$ (from the persistent Fe~K emission, which has been radially-integrated across the
{\it entire} emitting region of the accretion disk).  We will deal with the effect of measurement
uncertainties in \thetaobs in \S\ref{uncertainties}.  Further suppose, for this example, that we
measure red and blue peak energies from the emission of an orbiting hotspot to be \gm$=0.22$ and
\gp$=1.04$.  Given these three variables (in practice, obtained from the data), we are then able to
construct \am versus $R$ curves for the measured \gm and \gp values.  
The units of distance to the hotspot are shown in \figavsr as distance from the horizon radius (\rdp),
as they are given in the transfer-function tables.  The solution for the black-hole spin and the
distance to the hotspot is found at the intersection of the \gm and \gp curves.  Therefore, for the
example shown in \figavsrp, the distance to the hotspot (from the horizon) is $\sim 1.3 \ R_{\rm G}$
and the spin is maximal (\am$\sim1$).
We overlay the boundary corresponding to \rms on \figavsr (dashed line).  
The region above this line
contains pairs of values 
(\amp, \rdp) that pertain to hotspot orbits only at
radii larger than \rms (see \S\ref{avsrresults}). 
For such hotspots with \rd$<4$\rgp,
the \rms boundary itself offers
constraints on black-hole spin.  These constraints improve
(in the sense of imposing a tighter range in \amp) as the orbital radius of the hotspot decreases.  

\begin{figure}[!htb]
\begin{center}
	\epsscale{1.0}
    	\plotone{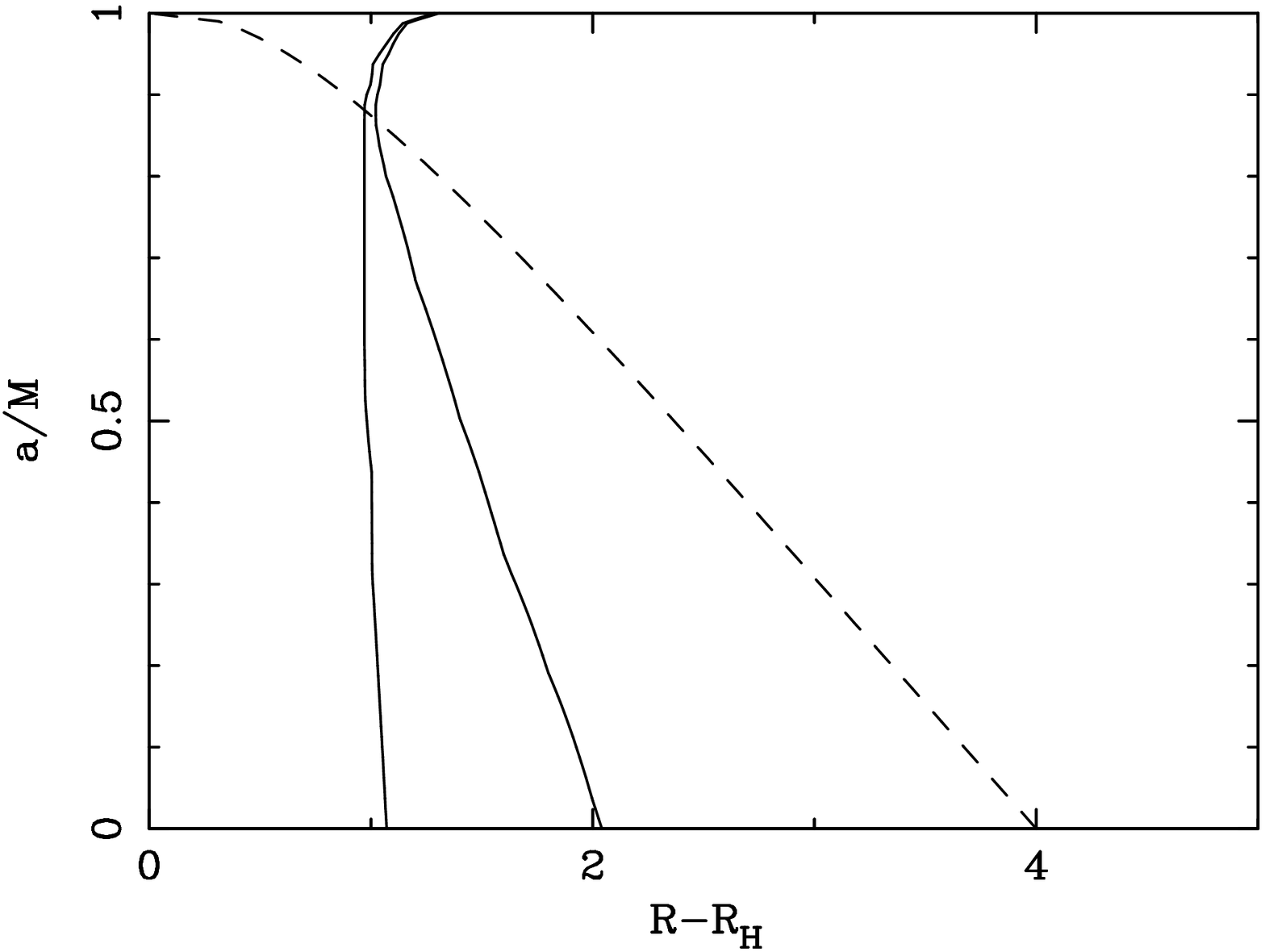}
\end{center}
\caption{\footnotesize Curves of \am versus distance to the hotspot from \rhorzp, the horizon radius,
(\rdp) for \gm$=0.22$ (left-hand, solid curve) and \gp$=1.04$ (right-hand, solid curve), 
with \thetaobs $=60^{\circ}$.  
The dashed line represents the \rms boundary. 
The solutions for the spin of the black hole and the 
distance to the hotspot are found at the intersection of the \gm and \gp curves.
Solutions that lie
above the \rms boundary curve correspond to orbits outside the ISCO ($R>R_{\rm MS}$).}
\end{figure}	

\subsection{Hotspot Fe~K Line Measurement Uncertainties}
\label{uncertainties}

The example shown in \figavsr assumes the ideal case of a full orbit of a point-like hotspot, and does
not include measurement uncertainties.  When applied to real data, there are observational
uncertainties that we must consider.  With future high-resolution, high-throughput instruments, the
largest uncertainty in the determination of \gm and \gp (after the finite size of the hotspot) will be
due to the unknown ionization state of Fe in the line emitting region.  Note that the local ionization
state in the vicinity of the hotspot may not necessarily be the same as the ionization state of the
entire accretion disk.  The rest energy of Fe~K line emission must be some value between  6.404 keV
(Fe~{\sc i}; see Bambynek 1972) and 6.966 keV (Fe~{\sc xxvi}; see Pike \etal 1996).  This corresponds
to an uncertainty of $\pm 4.2$ \% in the measurement of $g$ for the average rest-energy value in this
range.   We note, however, that the true error on $g$ due to an incorrectly assumed value of the
ionization state of Fe could be as large as twice this value.  Shown in \figavsrerr are the same \gm
and \gp curves as in \figavsrp, overlaid with curves corresponding to this $\pm 4.2$\% uncertainty.  Given
this uncertainty, the constraints on the black-hole spin and the distance to the hotspot for the
measured \thetaobsp, \gmp, and \gp are found in the overlap region of the two contours, shown in
\figavsrerrp.  

\begin{figure}[!htb]
\begin{center}
       \epsscale{1.0}
       \plotone{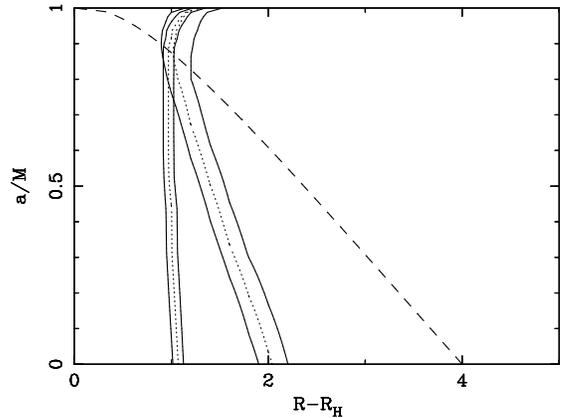}
\end{center}
\caption{\footnotesize Curves of \am versus distance to the hotspot from the horizon radius
(\rdp).  The dotted curves are the same as those in \figavsrp.  These curves are overlaid with
$\pm 4.2$\% error boundaries for \gm (left-hand contour) and \gp (right-hand contour), 
corresponding to the uncertainty in the 
ionization state of Fe.}
\end{figure}

To our knowledge, no method has yet been proposed for constraining the ionization state of Fe for a
localized region of the accretion disk.  If, in the future, it becomes possible to independently
measure this ionization state (e.g. by using the \fekb line in tandem with the \feka line), then
the largest uncertainty in actually measuring \gm and \gp will be due to the absolute energy scale
of the instrument.  For future instruments, this could be as small as 1 eV, corresponding to an
error of less than 0.02\%, and we could therefore obtain much narrower \gm and \gp contours.  A
separate source of error arises from uncertainties in the measurement of the disk inclination angle
obtained from the radially-integrated disk emission line.  It can be expected that future
instruments will yield \thetaobs measurements with uncertainties better than $\pm 1^{\circ}$.  Some
current data already yield \thetaobs uncertainties of this order (e.g. Brenneman \& Reynolds
2006).  \figthetaerrp, for example, shows the $a/M$ versus $R-$\rhorz curves for \gp$=0.5$, for
$29^{\circ}, 30^{\circ}, \ \rm and \ 31^{\circ}$.  We note that the error in \thetaobs produces a
much less significant effect in the \gm and \gp curves than that of the ionization uncertainty. 
However, the finite size of a hotspot will effectively give a finite width to the \gm and \gp
curves.  If there is significant line emission away from the center of the hotspot, the effect on
the width of the \gm and \gp curves {\it could} be worse than that corresponding to the ionization
uncertainty.  The unknown angular distribution of photon emission from the hotspot could also yield
greater uncertainty than that from the ionization state of Fe.  For example, if the emission from the
hotspot is highly anisotropic, the peak values measured from the data could correspond to values of $g$
other than the \gm and \gp values corresponding to the case of isotropic emission.

\begin{figure*}[!htb]
\begin{center}
	\epsscale{1.2}
        \plotone{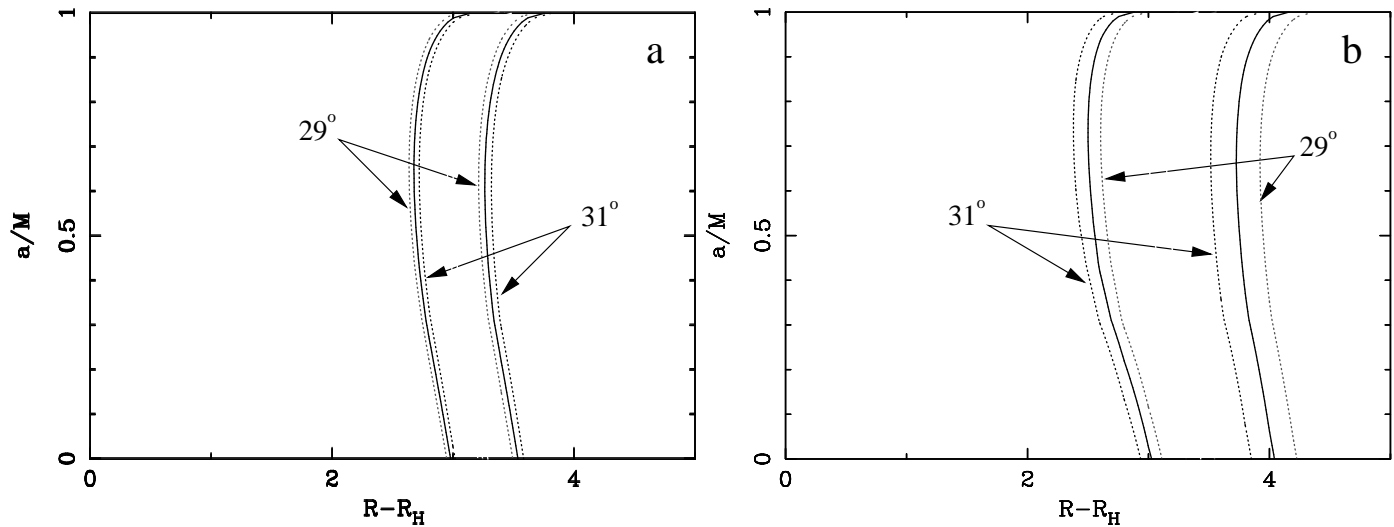}
\end{center}
\caption{\footnotesize 
	Uncertainty boundaries for $\pm 4.2$\% for (a) \gm$=0.519$ and (b) \gp$=0.892$, for 
       \thetaobs $=29^{\circ} \rm (dotted), \ 30^{\circ} (solid), \ and
       \ 31^{\circ} (dotted)$.  The effect of the $\pm 4.2$\% uncertainty in the rest energy
       of the Fe~K line due to the unknown ionization
       state is much greater than the 
       $\pm 1^{\circ}$ uncertainty in \thetaobs measurements.}
\end{figure*}

\section{Constraining Black-Hole Spin} \label{avsrresults} It remains unclear whether the emission from within the ISCO
contributes significantly to Fe~K line emission from the accretion
disk.  
Strictly speaking, a point-like hotspot could spiral past the 
ISCO, executing another orbit or more.  In this
case, radiation from the hotspot from within the ISCO could potentially contribute to the Fe K line
profile, even if it is only a restricted transition region (e.g. see Reynolds \& Fabian 2008). However, for 
a spirally-infalling hotspot the \gm and/or \gp peaks become
distorted and broadened (e.g. see Hartnoll \& Blackman, 2002; Fukumura \& Tsuruta, 2004). In such cases, in
practice, one would simply reject the data and search for another hotspot that had a cleaner signature.
Thus, if we have a hotspot producing clearly-defined \gm and/or \gp peaks, the \rms boundary gives a rough
constraint on \am and \rd for hotspots orbiting close to the black hole (\rd$\equiv R-R_{\rm H}\le 4$).  In
fact, this constraint is better than any constraint we could derive from the method described above, given
the $\pm4.2$\% error from the uncertainty in the ionization state of Fe, as is clear in the example in
\figavsrerrp.  We have also pointed out that additional uncertainties could result from the finite size of
the hotspot and from the departure from isotropic emission in the disk frame.  Furthermore, in the small
\rd regime, measurement of just {\it one} peak (\gm {\it or} \gpp) will yield such constraints from the
\rms boundary alone.  Naturally, if the hotspot is orbiting at $R>6$\rg (corresponding to \rd$>4$\rg for
\am$=0$ and \rd$>5$\rg for \am$=1$), the \rms boundary does not constrain spin.

\begin{figure*}[!tbh]
\begin{center}
	\epsscale{1.0}
    	\plotone{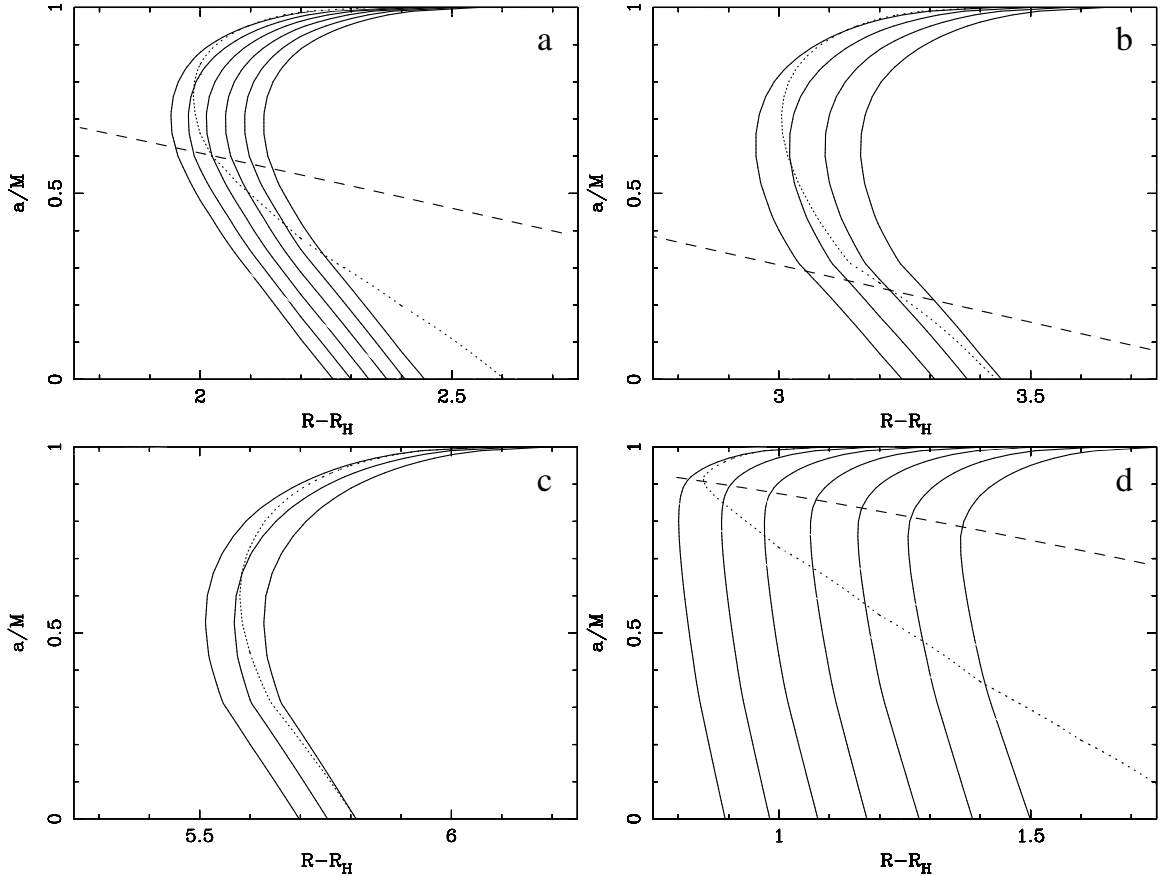}
\end{center}
	\caption{\footnotesize Curves 
	of \am versus \rd for various values of \gm (solid) and \gp ((a) 0.802, (b) 0.0892, and (c,d) 0.983; 
	dotted), for
	\thetaobs$=30^{\circ}$ in panels~a--c, 
	and $60^{\circ}$ in panel~d (see \S\ref{avsrresults} for values of \gm and further details).  
	Different pairs of (\gmp,\gpp) yield different solutions for the spin
	of the black hole and the distance to the orbiting hotspot.  The overlaid, dashed 
	line represents the \rms boundary.  Possible solutions occur at the intersection of
	the \gm and \gp curves. Solutions above the \rms boundary curve correspond to hotspot
	orbits that lie outside the ISCO ($R>R_{\rm MS}$).
	In (c), all intersections are solutions that lie outside the ISCO.}
\end{figure*}

In \figavsrexamps we show examples of \am versus \rd curves for possible measured values of \gp (dotted), as
well as curves for possible corresponding values of \gm (solid) for each \gpp.  Overlaid on each of these
plots, where applicable, is the curve marking the boundary of \rms (dashed line), as discussed in
\S\ref{hotspotmeth}.  In \figavsrexampsp a, the dotted curve corresponds to a measurement of \gp$=0.802$,
for \thetaobs$=30^{\circ}$; this is overlaid with (solid) curves corresponding to \gm values ranging from
0.426 to 0.446, with $\Delta$\gm=0.004 (i.e. from left to right, \gm has values of  0.426, 0.430, 0.434,
0.438, 0.442, 0.446).  These values clearly give distinct solutions for the value of \amp.  
Note that 
if we were only considering hotspots that complete full orbits outside of the ISCO ($R>R_{\rm MS}$), then
only the first three values of \gm
would be acceptable solutions for the given \gp value.
In practice, the net effect of all the measurement uncertainties discussed in \S\ref{uncertainties}
must produce finite widths in the \gm and \gp  curves
that are small enough to achieve the 
desired constraints on $a/M$. The $\Delta g_{-}$ intervals shown in
the examples in \figavsrexamps serve as a guide to what might be required (i.e.
net uncertainties in \gm and \gp better than $\sim 1\%$).
\figavsrexampssp b-d
show further examples of \am versus \rd contour plots.  In \figavsrexampsp b, we show results for values of
\gp$=0.892$ and \gm$=0.519 \ \rm to \ 0.534$ with $\Delta$\gm$=0.005$ for \thetaobs$=30^{\circ}$.
A plot for \gp$=0.983$, overlaid with \gm$=0.649$, 0.651, and 0.653 for \thetaobs$=30^{\circ}$ is shown in
\figavsrexampsp c. This is an example of a hotspot orbiting in the regime where the \rms
boundary cannot constrain the spin.  As \thetaobs increases, \gm and \gp are affected by larger Doppler
shifts.  In \figavsrexampsp d we show the plot corresponding to the same \gp value as in \figavsrexampsp c
(\gp$=0.983$), but this time for a different inclination angle, \thetaobs$=60^{\circ}$.  The solid curves
correspond to \gm$=0.190 \rm \ to \ 0.280$, with $\Delta$\gm$=0.015$.

We point out that the \gm and \gp functions are more sensitive to the spin when the hotspot is closer to
the black hole.  We see in \figavsrexampssp a-c that, as the orbiting radius increases, the
intersection of the \gm and \gp contours becomes increasingly difficult to discern, making it more
difficult to constrain \am once measurement errors have been taken into account.  However, the plots
in \figavsrexamps also show that the technique described here has the {\it potential} to constrain
the black-hole spin using Fe~K line profiles from orbiting hotspots, under favorable circumstances. 

\section{Conclusions} 
\label{conclusions} 

We have described a technique to measure black-hole spin based on
hotspots of enhanced Fe~K line emission co-rotating in an
accretion disk and studied the feasibility of the method.
An advantage of the hotspot method is that it does not
require knowledge of the Fe~K line radial emissivity function over the
disk. The method does not rely on measurement of line intensities,
only peak energies (\gm and \gpp) of narrow spectral features.
However, achieving sufficiently small measurement
uncertainties will be challenging and we have attempted to
quantify the principal sources of uncertainty. 

One caveat is that we assume that an independent measurement of the 
disk inclination angle,
\thetaobsp, has been made from spectral fitting of the persistent,
time-averaged, radially-integrated Fe~K emission.  However,
the measurement of \thetaobsp, which is most sensitive to the blue cut-off of the line emission, is much
less dependent on the other parameters of the model than spin, although the derived \thetaobs is subject to
some uncertainty due to the unknown ionization state of Fe.  
Another caveat is that departure from the assumptions of a standard,
geometrically-thin, optically-thick, Keplerian disk may introduce additional uncertainties in the
determination of black-hole spin.  

The technique described here {\it could}, in theory, be applied to real data obtained with future
high-spectral-resolution and high-throughput X-ray instrumentation.
In the present work we focused on the possibility of
constraining spin from {\it one} Fe~K emission line, most likely \fekap, due to a hotspot.  We note that it
may be possible to detect both \feka {\it and} \fekb emission, in which case we would have four separate
contours to constrain \am and \rd and we would furthermore be able to constrain the ionization state in the
vicinity of the hotspot to lower than Fe~{\sc xvii}, as \fekb cannot be produced for higher ionization
states.  Since the \feka line energy for Fe~{\sc xvii} is 6.43 keV (e.g. Mendoza \etal 2004), the
measurement uncertainties in \gm and \gp could be as low as 0.5\%.  
An additional possibility in the longer-term future
is to measure the orbital radius using
micro-arcsecond X-ray imaging instruments. This would
facilitate remarkable constraints on black-hole spin
from hotspots.

It is also important to consider that these results assume observations of a {\it complete orbit} of a
hotspot and, if it is unknown whether a full or partial orbit has been observed, there will be some
ambiguity in the observed double-horned Fe~K lines from hotspot observations.  For this reason, this
type of analysis would be best done with time-resolved spectroscopy, since the extreme redshift and
blueshift could be more easily verified.  Time-resolved spectroscopy of hotspots, under favorable
conditions, may be possible with micro-calorimeters onboard planned missions.  These instruments
promise a spectral resolution of a few eV in the Fe~K band combined with large effective area. Such
X-ray spectroscopic instruments will not only be able to resolve these double-horned profiles, but will
also yield {\it time-resolved} data that will allow us to break down the profile into multiple segments
as the hotspot orbits the black hole.   Therefore, with instrumentation on future missions, it may in
principle be possible to ``watch" an orbit with Fe~K line profiles that incrementally redshift (and
blueshift), tracing out the full, time-averaged, double-peaked profile.  It would then be possible to
constrain the orbital time of the hotspot.  Given that $t_{\rm orbit}$ depends on $M$, \amp, and
distance from the black hole to the hotspot, its measurement, in addition to the constraints on \rd
obtained by the independent measurement of \gm and \gp (as described here) may offer stronger
constraints on the spin.  This technique, however, could require more precise black-hole mass
measurements than those that are currently available.  We defer an in-depth investigation of the
feasibility of measuring spin with {\it time-resolved} X-ray spectroscopy to future work.

\begin{acknowledgements}
KM and TY acknowledge partial support from NASA grant NNG04GB78A.  VK and MD thank U.S.-Czech 
S\&T cooperation project ME09036.  KM and TY thank J. H. Krolik, J. D. Schnittman, and S. C.
Noble for discussions and comments. 

\end{acknowledgements}

{}
\end{document}